# Densest Subgraph in Streaming and MapReduce

Bahman Bahmani
Stanford University
Stanford, CA
bahman@stanford.edu*

Ravi Kumar
Yahoo! Research
Sunnyvale, CA
ravikumar@yahoo-inc.com

Sergei Vassilvitskii
Yahoo! Research
New York, NY
sergei@yahoo-inc.com

## ABSTRACT

The problem of finding locally dense components of a graph is an important primitive in data analysis, with wide-ranging applications from community mining to spam detection and the discovery of biological network modules. In this paper we present new algorithms for finding the densest subgraph in the streaming model. For any $\epsilon > 0$, our algorithms make $O(\log_{1+\epsilon} n)$ passes over the input and find a subgraph whose density is guaranteed to be within a factor $2(1+\epsilon)$ of the optimum. Our algorithms are also easily parallelizable and we illustrate this by realizing them in the MapReduce model. In addition we perform extensive experimental evaluation on massive real-world graphs showing the performance and scalability of our algorithms in practice.

## 1. INTRODUCTION

Large-scale graph processing remains a challenging problem in data analysis. In this work we focus on the densest subgraph problem that forms a basic primitive for a diverse number of applications ranging from those in computational biology [36] to community mining [8, 17] and spam detection [21]. We present algorithms that work both in the data streaming and distributed computing models for large scale data analysis and are efficient enough to generalize to graphs with billions of nodes and tens of billions of edges.

As input to the densest subgraph problem, we are given a graph $G = (V, E)$ and are asked to find a subset $S$ of nodes that has the highest ratio of edges between pairs of nodes in $S$ to the nodes in $S$. This basic problem can take on several flavors. The graph may be undirected (e.g. friendships in Facebook) or directed (e.g. followers in Twitter). In the latter case, the goal is to select two subsets $S$ and $T$ maximizing the number of edges from $S$ to $T$ normalized by the geometric mean of $|S|$ and $|T|$. A different line of work insists that the subgraphs be large: the input is augmented with an integer $k$ with the requirement that the output subset has at least $k$ nodes.

This simple problem has a variety of applications across different areas. We illustrate some examples below.

*(1) Community mining.* One of the most natural applications of the densest subgraph problem is finding structure in large networks. The densest subgraph problem is useful in identifying communities [12, 17, 32], which can then be leveraged to obtain better graph compression [8]. Heuristics, with no provable performance guarantees, have been typically used in this line of work.

*(2) Computational biology.* Saha et al. [36] adapt the densest subgraph problem for finding complex patterns in the gene annotation graph, using approximation and flow-based exact algorithms. They validate this approach by showing that some of the patterns automatically discovered had been previously studied in the literature; for more examples, see [1, Chapter 18].

*(3) Link spam detection.* Gibson et al. [21] observe that dense subgraphs on the web often correspond to link spam, hence their detection presents a useful feature for search engine ranking; they use a heuristic method that works well in the data stream paradigm.

*(4) Reachability and distance query indexing.* Algorithms for the densest subgraph problem form a crucial primitive in the construction of efficient indexes for reachability and distance queries, most notably in the well-known 2-hop labeling, first introduced in [14], as well as the more recent 3-hop indexing [23]. To underscore the importance of practical algorithms the authors of [14] remark that the 2-approximation algorithm of [10] is of more practical interest than the more complex but exact algorithm.

In all these applications, a good approximation to the densest subgraph is sufficient and is certainly more desirable than a heuristic without any performance guarantees.

It is known that both the directed and the undirected version of the densest subgraph problem can be solved optimally using parametric flow [29] or linear programming relaxation [10]. In the same work, Charikar [10] gave simple combinatorial approximation algorithms for this problem. On a high level, his algorithm for the undirected case greedily removes the worst node from the graph in every pass; the analysis shows that one of the intermediate graphs is a 2-approximation to the densest subgraph problem. The basic version of the problem provides no control over the size of the densest subgraph. But, if one insists on finding a large dense subgraph containing at least $k$ nodes, the problem be-

*Research supported in part by William R. Hewlett Stanford Graduate Fellowship, and NSF awards 0915040 and IIS-0904325. Part of this work was done while the author was at Yahoo! Research.

comes NP-hard [26]. Andersen and Chellapilla [3] as well as Khuller and Saha [26] show how to obtain 2-approximations for this version of the problem.

While the algorithms proposed in the above line of work guarantee good approximation factors, they are not efficient when run on very large datasets. In this work we show how to use the principles underlying existing algorithms, especially [10], to develop new algorithms that can be run in the data stream and distributed computing models, for example, MapReduce; this also resolves the open problem posed in [1, 13].

## 1.1 Streaming and MapReduce

As the datasets have grown to tera- and petabyte input sizes, two paradigms have emerged for developing algorithms that scale to such large inputs: streaming and MapReduce.

In the streaming model [34], one assumes that the input can be read sequentially in a number of passes over the data, while the total amount of random access memory (RAM) available to the computation is sublinear in the size of the input. The goal is to reduce the number of passes needed, all the while minimizing the amount of RAM necessary to store intermediate results. In the case the input is a graph, the nodes $V$ are known in advance, and the edges are streamed (it is known that most non-trivial graph problems require $\Omega(|V|)$ RAM, even if multiple passes can be used [18]). The challenge in streaming algorithms lies in wisely using the limited amount of information that can be stored between passes.

Complementing streaming algorithms, MapReduce, and its open source implementation, Hadoop, has become the de-facto model for distributed computation on a massive scale. Unlike streaming, where a single machine eventually sees the whole dataset, in MapReduce, the input is partitioned across a set of machines, each of which can perform a series of computations on its local slice of the data. The process can then be repeated, yielding a multi-pass algorithm (See [16] for exact framework, and [19, 25] for theoretical models). It is well known that simple operations like sum and other holistic measures [35] as well as some graph primitives, like finding connected components [25], can be implemented in MapReduce in a work-efficient manner. The challenge lies in reducing the total number of passes with no machine ever seeing the entire dataset.

## 1.2 Our contributions

In this work we focus on obtaining efficient algorithms for the densest subgraph problem that can work on massive graphs, where the graph cannot be stored in the main memory.

Specifically, we show how to modify the approach of [10] so that the resulting algorithm makes only $O(\frac{1}{\epsilon} \log n)$ passes over the data and guarantees to return an answer within a $(2+\epsilon)$ factor of optimum. We show that our algorithm only requires the computation of basic graph parameters (e.g., the degree of each node and the overall density) and thus can be easily parallelized—we use the MapReduce model to demonstrate one such parallel implementation. Finally, we show that despite the $(2+\epsilon)$ worst-case approximation guarantee, the algorithm's output is often nearly optimal on real-world graphs; moreover it can easily scale to graphs with billions of edges.

## 2. RELATED WORK

The densest subgraph problem lies at the core of large scale data mining and as such it and its variants have been intensively studied. Goldberg [22] was one of the first to formally introduce the problem of finding the densest subgraph in an undirected graph and gave an algorithm that required $O(\log n)$ flow computations to find the optimal solution; see also [29]. Charikar [10] described a simple greedy algorithm and showed that it leads to a 2-approximation to the optimum. When augmented with a constraint requiring the solution be of size at least $k$, the problem becomes NP-hard [26]. On the positive side, Andersen and Chellapilla [3] gave a 2-approximation to this version of the problem, and [26] gave a faster algorithm that achieves the same solution quality.

In the case the underlying graph is directed, Kannan and Vinay [24] were the first to define the notion of density and gave an $O(\log n)$ approximation algorithm. This was further improved by Charikar [10] who showed that it can be solved exactly in polynomial time by solving $O(n^2)$ linear programs, and obtained a combinatorial 2-approximation algorithm. The latter algorithm was simplified in the work of Khuller and Saha [26].

In addition to the steady theoretical progress, there is a rich line of work that tailored the problem to the specific task at hand. Variants of densest subgraph problem have been used in computational biology (see, for example [1, Chapter 14]), community mining [12, 17, 32], and even to decide what subset of people would form the most effective working group [20]. The specific problem of finding dense subgraphs on very large datasets was addressed in Gibson et al. [21] who eschewed approximation guarantees and used shingling approaches to find sets of nodes with high neighborhood overlap.

**Streaming and MapReduce.** Data streaming and MapReduce have emerged as two leading paradigms for handling computation on very large datasets. In the data stream model, the input is assumed too large to fit into main memory, and is instead streamed past one object at a time. For an introduction to streaming, see the excellent survey by Muthukrishnan [34]. When streaming graphs, the typical assumption is that the set of nodes is known ahead of time and can fit into main memory, and the edges arrive one by one; this is the semi-streaming model of computation [18]. Algorithms for a variety of graph primitives from matchings [31], to counting triangles [5, 6] have been proposed and analyzed in this setting.

While data streams are an efficient model of computation for a single machine, MapReduce has become a popular method for large-scale parallel processing. Beginning with the original work of Dean and Ghemawat [16], several algorithms have been proposed for distributed data analysis, from clustering [15] to solving set cover [13]. For graph problems, Karloff et al. [25] give algorithms for finding connected components and spanning trees; Suri and Vassilvitskii show how to count triangles effectively [37], while Lattanzi et al. [28] and Morales et al. [33] describe algorithms for finding matchings on massive graphs.

## 3. PRELIMINARIES

Let $G = (V, E)$ be an undirected graph. For a subset $S \subseteq V$, let the *induced edge set* be defined as $E(S) = E \cap S^2$

and let the *induced degree* of a node $i \in S$ be defined as $\deg_S(i) = |\{j \mid (i,j) \in E(S)\}|$.

The following notion of graph density is classical (see, for example, [29, Chapter 4]).

DEFINITION 1 (DENSITY, UNDIRECTED). *Let $G = (V, E)$ be an undirected graph. Given $S \subseteq V$, its* density $\rho(S)$ *is defined as*

$$\rho(S) = \frac{|E(S)|}{|S|}.$$

*The maximum density $\rho^*(G)$ of the graph is then*

$$\rho^*(G) = \max_{S \subseteq V} \{\rho(S)\}.$$

In case the graph is weighted, the density incorporates the total weight of all of the edges in the induced subgraph, $\rho(S) = \frac{\sum_{e \in E(S)} w_e}{|S|}$.

We also define density above a size threshold: given $k > 0$ and an undirected graph $G$, we define

$$\rho^*_{\geq k}(G) = \max_{S \subseteq V, |S| \geq k} \rho(S).$$

For directed graphs, the density is defined as follows [24]. Let $G = (V, E)$ be a directed graph. For $S, T \subseteq V$, where the subsets are not necessarily disjoint, let $E(S, T) = E \cap (S \times T)$. We abbreviate $E(\{i\}, T)$ as $E(i, T)$ and $E(S, \{j\})$ as $E(S, j)$.

DEFINITION 2 (DENSITY, DIRECTED). *Let $G = (V, E)$ be a directed graph. Given $S, T \subseteq V$, their* density $\rho(S, T)$ *is defined as*

$$\rho(S, T) = \frac{|E(S, T)|}{\sqrt{|S||T|}}.$$

*The maximum density $\rho^*(G)$ of the graph is then*

$$\rho^*(G) = \max_{S, T \subseteq V} \{\rho(S, T)\}.$$

**Approximation.** For $\alpha \geq 1$, an algorithm is said to obtain an $\alpha$-*approximation* to the undirected densest subgraph problem if it outputs a subset $\tilde{S} \subseteq V$ such that $\rho(\tilde{S}) \geq \rho^*(G)/\alpha$. An analogous definition can be made for the directed case.

## 4. ALGORITHMS

In this section we present streaming algorithms for finding approximately densest subgraphs. For any $\epsilon > 0$, we obtain a $(2 + 2\epsilon)$-approximation algorithm for the case of both undirected graphs (Section 4.1) and directed graphs (Section 4.3) and a $(3 + 3\epsilon)$-approximation algorithm when the densest subgraph is prescribed to be more than a certain size (Section 4.2). All of our algorithms make $O(\log n)$ passes over the input graph and use $O(n)$ main memory.

Our algorithms are motivated by Charikar's greedy algorithm for the densest subgraph problem [10] and the MapReduce algorithm for maximum coverage [7, 13]. They work by carefully relaxing the greedy constraint in a way that almost preserves the approximation factor, yet exponentially decreases the number of passes. We also show a lower bound on the space required by any streaming algorithm to obtain a constant-factor approximation.

### 4.1 Undirected graphs

In this section we present the greedy approximation algorithm for undirected graphs. Let $G = (V, E)$ be an undirected graph and let $\epsilon > 0$. The algorithm proceeds in passes, in every pass removing a constant fraction of the remaining nodes. We show that one of the intermediate subgraphs forms a $(2 + 2\epsilon)$-approximation to the densest subgraph. We note that the densest subgraph problem in undirected graphs can be solved exactly in polynomial time via flows or linear programming (LPs); however flow and LP techniques scale poorly to internet-sized graphs. We will show in Section 6 that despite worst-case examples, the algorithms we give yield near-optimal solutions on real-world graphs and are much simpler and more efficient than the flow/LP-based algorithms.

Starting with the given graph $G$, the algorithm computes the current density, $\rho(G)$, and removes all of the nodes (and their incident edges) whose degree is less than $(2+2\epsilon) \cdot \rho(G)$. If the resulting graph is non-empty, then the algorithm recurses on the remaining graph, with node set denoted by $S$, again computing its density and removing all of the nodes whose degree is lower than the specified threshold; we denote these nodes by $A(S)$. Then, the node set reduces to $S \setminus A(S)$, and the recursion continues in the same way. Algorithm 1 presents the complete description.

---

**Algorithm 1** Densest subgraph for undirected graphs.
---
**Require:** $G = (V, E)$ and $\epsilon > 0$
1: $\tilde{S}, S \leftarrow V$
2: **while** $S \neq \emptyset$ **do**
3:     $A(S) \leftarrow \{i \in S \mid \deg_S(i) \leq 2(1+\epsilon)\rho(S)\}$
4:     $S \leftarrow S \setminus A(S)$
5:     **if** $\rho(S) > \rho(\tilde{S})$ **then**
6:         $\tilde{S} \leftarrow S$
7:     **end if**
8: **end while**
9: **return** $\tilde{S}$

---

Clearly, this algorithm can be implemented in a streaming fashion using only $O(n)$ memory since we only need to store and update the current node degrees to compute the density and to decide which nodes to remove. We now analyze the approximation factor of the algorithm and its running time.

LEMMA 3. *Algorithm 1 obtains a $(2 + 2\epsilon)$-approximation to the densest subgraph problem.*

PROOF. As the algorithm proceeds, the density of the remaining graph is non-monotonic, a fact that we observe experimentally in Section 6. We will show, however, that one of the intermediate subgraphs is a $(2+2\epsilon)$-approximation to the optimal solution.

To proceed, fix some optimal solution $S^*$, s.t. $\rho(S^*) = \rho^*(G)$. First, we note that for each $i \in S^*$, $\deg_{S^*}(i) \geq \rho(S^*)$: indeed, by the optimality of $S^*$, for any $i \in S^*$, we have

$$\frac{|E(S^*)|}{|S^*|} = \rho(S^*) \geq \rho(S^* \setminus \{i\}) = \frac{|E(S^*)| - \deg_{S^*}(i)}{|S^*| - 1}. \quad (4.1)$$

Since $\sum_{i \in S} \deg_S(i) = 2|S|\rho(S)$, at least one node must be removed in every pass. Now, consider the first time in the pass when a node $i$ from the optimal solution $S^*$ is removed, i.e. $A(S) \cap S^* \neq \emptyset$; this moment is guaranteed to exist, since $S$ eventually becomes empty. Clearly, $S \supseteq S^*$. Let $i \in A(S) \cap S^*$. We have

$$\begin{aligned}
\rho(S^*) &\leq \deg_{S^*}(i) & \because (4.1) \\
&\leq \deg_S(i) & \because S \supseteq S^* \\
&\leq (2+2\epsilon)\rho(S). & \because i \in A(S)
\end{aligned}$$

This implies $\rho(S) \geq \rho(S^*)/(2+2\epsilon)$ and hence the algorithm outputs a $(2+2\epsilon)$-approximation. □

Next, we show that the algorithm removes a constant fraction of all of the nodes in every pass, and thus is guaranteed to terminate after $O(\log n)$ passes of the **while** loop.

LEMMA 4. *Algorithm 1 terminates in $O(\log_{1+\epsilon} n)$ passes.*

PROOF. At each step of the pass, we have

$$\begin{aligned}
2|E(S)| &= \sum_{i \in A(S)} \deg_S(i) + \sum_{i \in S \setminus A(S)} \deg_S(i) \\
&> 2(1+\epsilon)(|S| - |A(S)|)\rho(S) \\
&= 2(1+\epsilon)(|S| - |A(S)|)\frac{|E(S)|}{|S|},
\end{aligned}$$

where the second inequality follows by considering only those nodes in $S \setminus A(S)$. Thus,

$$|A(S)| > \frac{\epsilon}{1+\epsilon}|S|. \qquad (4.2)$$

Equivalently,

$$|S \setminus A(S)| < \frac{1}{1+\epsilon}|S|.$$

Therefore, the cardinality of the remaining set $S$ decreases by a factor at least $1/(1+\epsilon)$ during each pass. Hence, the algorithm terminates in $O(\log_{1+\epsilon} n)$ passes. □

Notice that for small $\epsilon$, $\log(1+\epsilon) \approx \epsilon$ and hence the number of passes is $O(\frac{1}{\epsilon} \log n)$.

### 4.1.1 Lower bounds

In this section we show that our analysis is tight. In particular, we show that there are graphs on which Algorithm 1 makes $\Omega(\log n)$ passes. Furthermore, we also show that any algorithm that achieves a 2-approximation in $O(\log n)$ passes must use $\Omega(n/\log n)$ space. Note that Algorithm 1 comes close to this lower bound since it makes $O(\log n)$ passes and uses $O(n)$ memory.

**Pass lower bound.** We show that the analysis of the number of passes is tight up to constant factors. We begin with a slightly weaker result.

LEMMA 5. *There exists an unweighted graph on which Algorithm 1 requires $\Omega(\frac{\log n}{\log \log n})$ passes.*

PROOF. The graph consists of $k$ disjoint subsets $G_1, \ldots, G_k$, where $G_i$ is a $2^{i-1}$ regular graph on $|V_i| = 2^{2k+1-i}$ nodes, hence every $G_i$ has exactly $2^{2k-1}$ edges and has density of $2^{i-2}$. For any $\ell \geq 1$, let $G_{\geq \ell} = \bigcup_{i \geq \ell} G_i$. The density $G_{\geq \ell}$ is:

$$\rho(G_{\geq \ell}) = \frac{(k-\ell+1)2^{2k-1}}{2^{k+1}(2^{k-\ell+1}-1)} \approx (k-\ell+1)2^{\ell-3}.$$

We claim that in every pass the algorithm removes $O(\log k)$ of these subgraphs. Suppose that we start with the subgraph $G_{\geq \ell}$ at the beginning of the pass. Then the nodes in $A(S)$ are exactly those that have their degree less than $\rho(G_{\geq \ell})(2+\epsilon) \approx (k-\ell+1)2^{\ell-2}$. Since a node in $G_i$ has degree $2^{i-1}$, this is equivalent to nodes in $G_i$ for $i < (\ell-1) + \log(k-\ell)$, and hence the subgraph in the next pass is $G_{\geq \ell+\log(k-\ell)-1}$.

Thus, the algorithm will take at least $\Omega(k/\log k)$ passes to complete. Since $k = \Theta(\log n)$, the proof follows. □

To show an example on which Algorithm 1 needs $\Omega(\log n)$ passes, we appeal to weighted graphs. Note that Algorithm 1 and the analysis easily generalize to finding the maximum density subgraph in an undirected weighted graph.

LEMMA 6. *There exists a weighted graph on which Algorithm 1 requires $\Omega(\log n)$ passes.*

PROOF SKETCH. Consider a graph whose degree sequence follows a power law with exponent $0 < \alpha < 1$, i.e., if $d_i$ is the $i$th largest degree, then $d_i \propto i^{-\alpha}$. We have $\sum_{i=1}^n i^{-\alpha} \simeq \int_0^n x^{-\alpha} dx = \frac{n^{1-\alpha}}{1-\alpha}$, so if the graph has $m$ edges, we (approximately) have $d_i = \frac{(1-\alpha)i^{-\alpha}m}{n^{1-\alpha}}$. Hence, in the first pass of the algorithm, we remove all the nodes with

$$d_i = \frac{(1-\alpha)i^{-\alpha}m}{n^{1-\alpha}} \leq (2+\epsilon)\frac{m}{n}.$$

Hence, the nodes such that

$$i < \left(\frac{1-\alpha}{2+\epsilon}\right)^{1/\alpha} n,$$

go to the next pass; note that this is a constant fraction of the nodes. As long as the power law property of the degree sequence is preserved after removing the low degree nodes in each pass, we obtain the desired $\Omega(\log n)$ lower bound.

Consider the graphs generated by the preferential attachment process [2]. To avoid the stochasticity in the model, which only makes the analysis more complicated, one can consider the following deterministic variant of this process: whenever a new node $u$ arrives, it adds an edge to all of the existing nodes $v$ and assigns a weight $w_{u,v}$ to the edge $(u,v)$ which is proportional to the current degree of $v$. Then degree of the $i$th node after a total of $n$ nodes have arrived follows a power law distribution which is exactly what we needed to achieve. □

**Space lower bound.** We show that the trade off between memory and number of passes is almost the best possible. Namely, any constant-pass streaming algorithm for approximating the densest subgraph to within a constant factor of 2 must use a linear amount of memory, and an algorithm making $O(\log n)$ passes must use $\Omega(n/\log n)$ memory.

LEMMA 7. *Any $p$-pass streaming $\alpha$-approximation algorithm for the densest subgraph problem, where $\alpha \geq 2$, needs $\Omega(n/(p\alpha^2))$ space.*

PROOF. Consider the standard disjointness problem in the $q$-party arbitrary round communication model. There are $q \geq 2$ players, and the $j$th player has the $n$-bit vector $x_{j1}, \ldots, x_{jn}$. Their goal is to decide if there is an index $i$ such that $\wedge_{j=1}^q x_{ji} = 1$. It is known that this problem needs $\Omega(n/q)$ communication [4, 9] and the lower bound holds even under the promise that either the bit vectors are pairwise disjoint (NO instance) or they have a unique common element but are otherwise disjoint (YES instance).

Given such an instance of disjointness, we construct the following densest subgraph instance. The overall graph $G = (V, E)$ consists of $n$ disjoint subgraphs $G_1, \ldots, G_n$. Each graph $G_i = (V_i, E_i)$ has $q$ nodes, $V_i = \{u_{1,i}, \ldots, u_{q,i}\}$. For each $i$, if $x_{ji} = 1$, then the $j$th player adds the $q-1$ edges $\{(u_{ji}, u_{j'i}) \mid j' \in [q], j' \neq j\}$ to $E_i$.

It is easy to see that, given the promise of a pairwise disjoint instance, each graph $G_i$ is a star in the case of a NO instance. In the case of a YES instance, one of the $G_i$'s is a complete graph. Therefore, $\rho(G) = (q-1)$ if and only if it is a YES instance and $\rho(G) = 1 - 1/q$ otherwise.

By setting $\alpha = q$ and using a standard reduction from streaming to communication we can conclude that if a $p$-pass streaming algorithm uses $o(n/(\alpha^2 p))$ memory and obtains an $\alpha$-approximation to the densest subgraph, then it can be used on $G$ to decide the disjointness instance, using $o(n)$ communication. Given the communication lower bound for disjointness, the space lower bound for the densest subgraph follows. □

## 4.2 Large dense subgraphs

In this section we show that a small modification of the algorithm presented in Section 4.1 gives a good approximation to the problem of finding densest subgraphs above a prescribed size, $k$.

The main difference from Algorithm 1 comes from the fact that instead of removing *all* of the nodes with degree less than $2(1+\epsilon)\rho(S)$, we only remove $\frac{\epsilon}{1+\epsilon}|S|$ of them. Intuitively, by removing the smallest number of nodes necessary to guarantee the fast convergence of the algorithm, we make sure that at least one of the graphs under consideration has approximately $k$ nodes. Algorithm 2 contains the complete description.

---

**Algorithm 2** Large densest subgraphs.

---
**Require:** $G = (V, E)$, $k > 0$, and $\epsilon > 0$
1: $S, \tilde{S} \leftarrow V$
2: **while** $S \neq \emptyset$ **do**
3:     $\widetilde{A}(S) \leftarrow \{i \in S \mid \deg_S(i) \leq 2(1+\epsilon)\rho(S)\}$
4:     Let $A(S) \subseteq \widetilde{A}(S)$, with $|A(S)| = \frac{\epsilon}{1+\epsilon}|S|$
5:     $S \leftarrow S \setminus A(S)$
6:     **if** $|S| \geq k$ and $\rho(S) > \rho(\tilde{S})$ **then**
7:         $\tilde{S} \leftarrow S$
8:     **end if**
9: **end while**
10: **return** $\tilde{S}$

---

To prove the approximation ratio of the algorithm, we use the following notation from [3].

DEFINITION 8. *The $d$-core of $G$, denoted $C_d(G)$, is the largest induced subgraph of $G$ with all degrees larger or equal to $d$.*

THEOREM 9. *Algorithm 2 is a $(3+3\epsilon)$-approximation algorithm for the problem of finding $\rho_{\geq k}(G)$.*

PROOF. The fact that $|\tilde{S}| \geq k$ is obvious from the definition of the algorithm. Let $S^* = \arg\max \rho_{\geq k}(G)$ and $\rho^* = \rho(S^*)$, and $\beta = \frac{2}{3(1+\epsilon)}$. Let $S$ be the first set generated during the algorithm such that $\rho(S) \geq \frac{\beta}{2}\rho^* = \frac{\rho^*}{3(1+\epsilon)}$. Such a set must exist, since $\rho_{\geq k}^* \leq \rho^*(G)$ and we saw in Lemma 3 that at least one of the generated sets has density at least $\frac{\rho^*}{2(1+\epsilon)}$. If $|S| \geq k$, then we are done.

Otherwise, consider the case when $|S| < k$. For any set $S'$ generated before $S$ during the algorithm, we have $\rho(S') < \frac{\beta}{2}\rho^*$. Then, for any node $i \notin S$, we have $d_i < 2(1+\epsilon)\frac{\beta}{2}\rho^* = (1+\epsilon)\beta\rho^*$, where $d_i$ is the degree of $i$ at the time it got removed during the algorithm. Thus, none of those nodes can be in the core $C_{(1+\epsilon)\beta\rho^*}(G)$. Therefore, $C_{(1+\epsilon)\beta\rho^*}(G) \subseteq S$, and:

$$
\begin{aligned}
|E(S)| &\geq |E(C_{(1+\epsilon)\beta\rho^*}(G))| && \because C_{(1+\epsilon)\beta\rho^*}(G) \subseteq S \\
&\geq |E(C_{(1+\epsilon)\beta\rho^*}(S^*))| \\
&\geq (1-(1+\epsilon)\beta)|E(S^*)|. && \because [3, \text{Lemma 2}]
\end{aligned}
\tag{4.3}
$$

Now, let $\widehat{S}$ be the last set generated during the algorithm such that $|\widehat{S}| \geq k$. We will show that $\rho(\widehat{S})$ is a $(3+3\epsilon)$-approximation to $\rho^*$. Since we remove at most $\frac{\epsilon}{1+\epsilon}|\widehat{S}|$ nodes, we have $k > |\widehat{S} - A(\widehat{S})| = \frac{|\widehat{S}|}{1+\epsilon}$, i.e., $|\widehat{S}| < (1+\epsilon)k$. Also, $S \subseteq \widehat{S}$, hence $|E(\widehat{S})| \geq |E(S)|$.

Therefore,

$$
\begin{aligned}
\frac{|E(\widehat{S})|}{|\widehat{S}|} &> \frac{|E(S)|}{(1+\epsilon)k} && \because S \subseteq \widehat{S}, |\widehat{S}| < (1+\epsilon)k \\
&\geq \frac{1-(1+\epsilon)\beta}{1+\epsilon}\rho^* && \because (4.3) \\
&= \frac{\beta}{2}\rho^* = \frac{1}{3(1+\epsilon)}\rho^*. \quad \square
\end{aligned}
$$

Although the algorithm above has a worse performance guarantee than Algorithm 1, this is only true in the case when the densest subgraph has fewer than $k$ nodes. In the case that the densest subgraph on $G$ has at least $k$ nodes then the above algorithm performs on par with Algorithm 1.

LEMMA 10. *Let $|S^*| > k$, where $S^* = \arg\max \rho_{\geq k}(G)$. Then, Algorithm 2 achieves a $(2+2\epsilon)$-approximation for $\rho_{\geq k}^*(G)$.*

PROOF. If $|S^*| > k$, and $\rho^* = \rho(S^*)$, then one can see that for any $i \in S^*$, $\deg_{S^*}(i) \geq \rho^*$. Now, consider the first set $S$ generated during the algorithm such that $A(S) \cap S^* \neq \emptyset$. Since the final set generated by the algorithm has cardinality $k$, and $|S^*| > k$, such a set definitely exists. For the considered set $S$, if $i \in A(S) \cap S^*$, then by the definition of $A(S)$, we have

$$2(1+\epsilon)\rho(S) \geq \deg_S(i) \geq \deg_{S^*}(i) \geq \rho^*,$$

completing the proof. □

Finally, to bound the number of passes, note that once the remaining subgraph has fewer than $k$ nodes, we can safely terminate the algorithm and return the best set seen so far. Together with Lemma 4 this immediately leads to the following.

LEMMA 11. *Algorithm 2 terminates in $O(\log_{1+\epsilon} \frac{n}{k})$ passes.*

## 4.3 Directed graphs

In this section we obtain a $(2+2\epsilon)$-approximation algorithm for finding the densest subgraph in directed graphs. Recall that in directed graphs we are looking for two not

458

necessarily disjoint subsets $S, T \subseteq V$. We assume that the ratio $c = |S^*|/|T^*|$ for the optimal sets $S^*, T^*$ is known to the algorithm. In practice, one can do a search for this value, by trying the algorithm for different values of $c$ and retaining the best result.

The algorithm then proceeds in a similar spirit as in the undirected case. We begin with $S = T = V$ and remove either those nodes $A(S)$ whose outdegree to $T$ is below average, or those nodes $B(T)$ whose indegree to $S$ is below average. (Formally we need the degrees to be below a threshold slightly above the average for the algorithm to converge.) A naive way to decide whether the set $A(S)$ or $B(T)$ should be removed in the current pass is to look at the maximum outdegree, $E(i^*, T)$, of nodes in $A(S)$ and the maximum indegree, $E(S, j^*)$, of nodes in $B(T)$. If $E(S, j^*)/E(i^*, T) \geq c$ then $A(S)$ can be removed and otherwise $B(T)$ can be removed. However, a better way is to make this choice directly based on the current sizes of $S$ and $T$. Intuitively, if $|S|/|T| > c$, then we should be removing the nodes from $S$ to get the ratio closer to $c$, otherwise we should remove those from $T$. In addition to being simpler, this way is also faster mainly due to the fact that it needs to compute *either* $A(S)$ or $B(T)$ in every pass, leading to a significant speedup in practice.

Algorithm 3 contains the formal description.

---

**Algorithm 3** Densest subgraph for directed graphs.

**Require:** $G = (V, E)$, $c > 0$, and $\epsilon > 0$
1: $\tilde{S}, \tilde{T}, S, T \leftarrow V$
2: **while** $S \neq \emptyset$ and $T \neq \emptyset$ **do**
3:    **if** $|S|/|T| \geq c$ **then**
4:       $A(S) \leftarrow \left\{ i \in S \mid |E(i, T)| \leq (1+\epsilon) \frac{|E(S,T)|}{|S|} \right\}$
5:       $S \leftarrow S \setminus A(S)$
6:    **else**
7:       $B(T) \leftarrow \left\{ j \in T \mid |E(S, j)| \leq (1+\epsilon) \frac{|E(S,T)|}{|T|} \right\}$
8:       $T \leftarrow T \setminus B(T)$
9:    **end if**
10:   **if** $\rho(S, T) > \rho(\tilde{S}, \tilde{T})$ **then**
11:      $\tilde{S} \leftarrow S, \tilde{T} \leftarrow T$
12:   **end if**
13: **end while**
14: **return** $\tilde{S}, \tilde{T}$

---

First, we analyze the approximation factor of the algorithm.

LEMMA 12. *Algorithm 3 leads to a $(2+2\epsilon)$-approximation to the densest subgraph problem on directed graphs.*

PROOF. As in [10], we generate an assignment of the edges to the endpoints corresponding to the algorithm. Whenever $(i, j) \in E$, and $i \in A(S)$ is removed from $S$, we assign $(i, j)$ to $i$; a similar assignment is made for the nodes in $B(T)$. Let $\tilde{\rho} = \rho(\tilde{S}, \tilde{T})$. Let $\deg^*_{\text{out}}$ be the maximum outdegree and $\deg^*_{\text{in}}$ be the maximum indegree in $G$.

We need to show that if $A(S)$ is removed, then

$$\forall i \in A(S), \sqrt{c}|E(i, T)| \leq (1 + \epsilon)\rho(S, T).$$

Suppose that $|S|/|T| \geq c$, and so the nodes in $A(S)$ will be removed. For all $i \in A(S)$, we have

$$\sqrt{c}|E(i, T)| \leq \sqrt{c} \cdot (1 + \epsilon) \frac{|E(S, T)|}{|S|}$$
$$\leq \sqrt{c} \cdot (1 + \epsilon)|E(S, T)| \sqrt{\frac{1}{c|S||T|}}$$
$$= (1 + \epsilon) \frac{|E(S, T)|}{\sqrt{|S||T|}}$$
$$= (1 + \epsilon)\rho(S, T).$$

The second line follows because $|S| \geq c|T| \Rightarrow |S| \geq \sqrt{c|S||T|}$. Similarly, one can show that if $B(T)$ gets removed, then

$$\forall j \in B(T), \frac{1}{\sqrt{c}} E(S, j) \leq (1 + \epsilon)\rho(S, T).$$

This proves that in the given assignment (of edges to endpoints), $\sqrt{c} \deg^*_{\text{out}} \leq (1+\epsilon)\tilde{\rho}$ and $\frac{1}{\sqrt{c}} \deg^*_{\text{in}} \leq (1+\epsilon)\tilde{\rho}$. Once we have such an assignment, we can use the same logic as in Lemmas 7 and 8 in [10] to conclude that the algorithm gives a $(2 + 2\epsilon)$-approximation:

$$\max_{S, T \subseteq V, |S|/|T|=c} \{\rho(S, T)\} \leq (2 + 2\epsilon)\rho(\tilde{S}, \tilde{T}). \quad \square$$

Next, we analyze the number of passes of the algorithm. The proof is similar to that of Lemma 4.

LEMMA 13. *Algorithm 3 terminate in $O(\log_{1+\epsilon} n)$ passes.*

PROOF. We have

$$|E(S, T)| = \sum_{i \in A(S)} |E(i, T)| + \sum_{i \in S \setminus A(S)} |E(i, T)|$$
$$> (1 + \epsilon)(|S| - |A(S)|) \frac{|E(S, T)|}{|S|},$$

which yields

$$|S \setminus A(S)| < \frac{1}{1 + \epsilon} |S|.$$

Similarly, we can prove

$$|T \setminus B(T)| < \frac{1}{1 + \epsilon} |T|.$$

Therefore, during each pass of the algorithm, either the size of the remaining set $S$ or the size of the remaining set $T$ goes down by a factor of at least $1/(1+\epsilon)$. Hence, in $O(\log_{1+\epsilon} n)$ passes, one of these sets becomes empty and the algorithm terminates. $\square$

## 5. PRACTICAL CONSIDERATIONS

In this section we describe two practical considerations in implementing the algorithms. The first (Section 5.1) is a heuristic method based on Count-Sketch to cut the memory requirements of the algorithm. The second (Section 5.2) is a discussion on how to realize the algorithms in the MapReduce computing model.

### 5.1 Heuristic improvements

We showed in Lemma 7 that any $p$-pass algorithm achieving a 2-approximation to the densest subgraph problem must use at least $\Omega(\frac{n}{p})$ space. However, even this amount of space can be prohibitively large for very large datasets. To further reduce the space required by the algorithms we turn to sketching techniques that probabilistically summarize the degree distribution of the nodes.

Recall that in order to decide whether to remove a particular node, the algorithm only needs to be aware of its degree. This is the same as counting the number of edges in the stream that share this node as one of their endpoints. This exact problem of maintaining the frequencies of items in the stream using sublinear space was addressed by Charikar et al. [11]. They introduce the *Count-Sketch* data structure, which maintains $t$ independent estimates, each as a table on $b$ buckets. For $i = 1, \ldots, t$, let $h_i : V \to [b]$ and $g_i : V \to \{\pm 1\}$ be hash functions and for $j = 1, \ldots, b$, let $c_{i,j}$ be counters initialized to zero. When an edge $(x, y)$ arrives, for each $i = 1, \ldots, t$, we update the counters as follows: $c_{i,h_i(x)} \leftarrow c_{i,h_i(x)} + g_i(x)$ and $c_{i,h_i(y)} \leftarrow c_{i,h_i(y)} + g_i(y)$. Finally, when queried for the final degree of a node $x$, we return the median among all of the estimates $\{c_{i,h_i(x)} \cdot g_i(x)\}_{i=1}^t$. (We refer the reader to [11] for the full details of the data structure; in our work, we merely use it as a black-box.)

Charikar et al. [11] showed that this way of probabilistically counting leads to a high precision counter for elements with high frequencies. Intuitively, this kind of a guarantee makes for a perfect fit with our application. We want to have good estimates for the nodes with high degrees (otherwise one may be removed prematurely). On the other hand, the false positive error of accidentally keeping a low-degree node is not as severe; a small number of low degree nodes will not have a dramatic impact on the size of the densest subgraph. As we show in Section 6 this intuition holds true in practice, and we find that the Count-Sketch enabled version of the algorithm, which uses a lot less space, sometimes (when lucky!) performs even better than the version using exact counting.

## 5.2 MapReduce implementation

All of the algorithms presented in this work depend on three basic functions: computing the density of the current graph, computing the degree of each individual node, and removing nodes with degree less than a specified threshold. The algorithm itself is very amenable to parallelism as long as these basic functions can exploit parallelism. For illustration purposes, we focus on a specific distributed computing model that is widely used in practice, namely, the MapReduce model. We assume familiarity with the MapReduce model of computation and refer the reader to [16] for details. Finding the best parallel implementation of our algorithm is an interesting future research direction.

Computing the density of the graph is a trivial operation, as one needs only to count the total number of edges and nodes present. To compute the degree of every node in parallel, in the map step duplicate each edge $(u, v)$ as two $\langle key; value \rangle$ pairs: $\langle u; v \rangle$ and $\langle v; u \rangle$. This way the input to every reduce task will be of the form $\langle u; v_1, v_2, \ldots, v_d \rangle$ where $v_1, v_2, \ldots, v_d$ are the neighbors of $u$ in $G$. The reducer can then count the number of associated values for each key, and output $\langle u; \deg(u) \rangle$.

Finally, the removal of the nodes with degree less than some threshold $t$, and their incident edges can be accomplished in two MapReduce passes. In the first map phase, we mark all of the nodes slated for removal by adding a $\langle v; \$\rangle$ key-value pair for all nodes $v$ that are being removed. We map each edge $(u, v)$ to $\langle u; v \rangle$. The reduce task associated with $u$ then gets all of the edges whose first endpoint is $u$, and the symbol $\$$ if the node was marked. In case the node is marked, the reduce task returns nothing, otherwise it just copies its input. In the second MapReduce pass we pivot on the second node in the edge description. Again, we only keep the edges incident on unmarked nodes. It is easy to see that the only edges that survive are exactly those incident on a pair of unmarked nodes.

## 6. EXPERIMENTS

In this section we detail the experiments and the results of the experiments for our algorithms. First, we describe the datasets used in our experiments (Section 6.1). These datasets are large social networks, some of which are publicly available for download or obtained through an API. Next, we study the accuracy of our algorithms when compared to the optimum. To this end, we obtain the optimum density value using a linear program, and compare the output of our algorithm to this optimum (Section 6.2). We then study the performance of the streaming version of our algorithm on both undirected and directed graphs. In particular, we analyze the effect of $\epsilon$ on the accuracy and the number of passes (Section 6.3 and Section 6.4). Finally, we remark (Section 6.5) on the space savings brought about by the sketching heuristic presented in Section 5.1 and on a proof-of-concept MapReduce implementation to compute the densest subgraph on an extremely large graph (Section 6.6).

Since we focus just on finding (or approximating) the densest subgraph, we do not try to enumerate *all* of the dense subgraphs in the given graph. It is easy to adapt our algorithm to iteratively enumerate node-disjoint (approximately) densest subgraphs in the graph, with the guarantee that at each step of the enumeration, the algorithm will produce an approximate solution on the residual graph. The quality of the resulting solution reflects more the properties of the underlying graph than our algorithm itself and hence we do not further explore this direction.

### 6.1 Data description

Almost all of our experiments are based on four large social networks, namely, FLICKR, IM, LIVEJOURNAL, and TWITTER. FLICKR is the social network corresponding to the Flickr (`flickr.com`) photosharing website, IM is the graph induced by the contacts in Yahoo! messenger service, LIVEJOURNAL is the graph induced by the friends in the LiveJournal (`livejournal.com`) social network, and TWITTER is the graph induced by the followers in the social media site Twitter (`twitter.com`).

FLICKR is available publicly and can be obtained by using an API (`www.flickr.com/services/api/`). A smaller version of the IM graph can be obtained via the Webscope program from `webscope.sandbox.yahoo.com/catalog.php?datatype=g`. The version of LIVEJOURNAL used in our experiments can be downloaded from `snap.stanford.edu/data/soc-LiveJournal1.txt.gz` and the version of TWITTER used in our experiments can be obtained from `an.kaist.ac.kr/~haewoon/release/twitter_social_graph/`. The details of the datasets are provided in Table 1.

Note that when trying to measure the quality of our algorithms, the following two baselines do not make sense in the context of the above graphs: (i) computing the actual densest subgraph, which is infeasible for such large graphs and (ii) running the algorithm of [10], which would take



| $G$ | type | $|V|$ | $|E|$ |
|---:|---|---|---|
| FLICKR | undirected | 976K | 7.6M |
| IM | undirected | 645M | 6.1B |
| LIVEJOURNAL | directed | 4.84M | 68.9M |
| TWITTER | directed | 50.7M | 2.7B |

Table 1: Parameters of the graphs used in the experiments.

quadratic time (linear time for each pass and a linear number of passes), which is still infeasible for these graphs. In order to circumvent this, we work with slightly smaller graphs just to compare the quality of the solution to that of the optimum (Section 6.2).

## 6.2 Quality of approximation

We study how good of an approximation is obtained by our algorithm for the undirected case. To enable this, we need to compute the value of the optimum. Recall that, as mentioned in section 1, both the directed and undirected densest subgraph problems can be solved exactly using parametric flow. In this section we want to obtain $\rho^*$, i.e., the value of the optimal solution, to argue that the approximation factor in practice is much better than $2(1+\epsilon)$, guaranteed by Lemma 3. (To do such a test for directed graphs is very expensive because one has to try all $n^2$ values of $c$.)

In order to solve the densest subgraph problem exactly, we use the following linear programming (LP) formulation.

$$\max \sum_{ij} x_{ij}$$
$$\forall (i,j) \in E, x_{ij} \leq y_i$$
$$\forall (i,j) \in E, x_{ij} \leq y_j$$
$$\sum_i y_i \leq 1$$
$$x_{ij}, y_i \geq 0$$

Charikar [10] showed that the value of this LP is precisely equal to $\rho^*(G)$. We use this observation to measure the quality of approximation obtained by Algorithm 1. To solve the LP, we use the COIN-OR CLP solver (`projects.coin-or.org/Clp`). We use seven moderately-sized undirected graphs publicly available at SNAP (`snap.stanford.edu`). Table 2 shows the parameters of these graphs and the approximation factor of our algorithms for different settings of $\epsilon$. It is clear that the approximation factors obtained by our algorithm are much better than what Lemma 3 promises. Furthermore, even high values of $\epsilon$ seem to hardly hurt the approximation guarantees.

## 6.3 Undirected graphs

In this section we study the performance of our algorithms on two undirected graphs, namely, FLICKR and IM. First, we study the effect of $\epsilon$ on the approximation factor and the number of passes. Figure 6.1 shows the results. For ease of comparison, we show the values relative to the density obtained by our algorithm for $\epsilon = 0$. (Note that the setting $\epsilon = 0$ is similar to Charikar's algorithm [10] in terms of the approximation factor but can run in much fewer number of passes; however, termination is not guaranteed for $\epsilon = 0$.) As we saw in Table 2, the approximation does not deteriorate for higher values of $\epsilon$ (note that the performance is not monotone in $\epsilon$). Choosing a value of $\epsilon \in [0.5, 1]$ seems to cut down the number of passes by half while losing only 10% of the optimum.

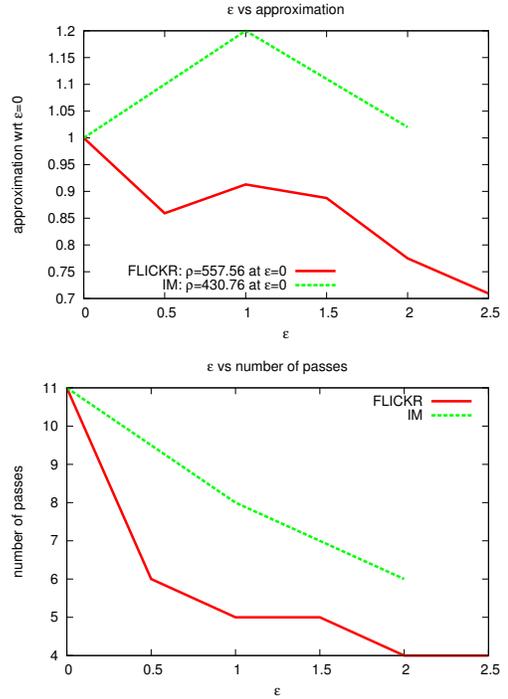

Figure 6.1: Effect of $\epsilon$ on the approximation and the number of passes.

We then move on to analyze the graph structure as the passes progress. Figure 6.2 shows the relative density as a function of the number of passes. (Curiously, we observe a unimodal behavior for FLICKR, but this does not seem to hold in general.)

Figure 6.3 shows the number of nodes and edges in the graph after each pass. The shape of the plots suggests that the graph gets dramatically smaller even in the early passes. This is a very useful feature in practice, since if the graph gets very small early on, then the rest of the computation can be done in the main memory. This will avoid the overhead of additional passes.

Note also that the worst-case bound of $O(\log_{1+\epsilon} n)$ for the number of passes as given by Lemma 4 is never achieved by these graphs. This is possibly because of the heavy-tail nature of the degree distribution of graphs derived from social networks and their core connectivity properties; see [27, 30]. These properties may also contribute to achieving the good approximation ratio, i.e., the worst-case bound of Lemma 3 is not met by these graphs. Exploring these in further detail is outside the scope of this work and is an interesting area of future research.

## 6.4 Directed graphs

In this section we study the performance of the directed graph version of our algorithm. We use the LIVEJOURNAL and TWITTER graphs for this purpose. Recall that for directed graphs, we have to try for various values of $c$ (Section 4.3). Of course, trying all $n^2$ possible values of $c$ is prohibitive. A simple alternative is to choose a resolution ($\delta >$

| $G = (V, E)$ | $|V|$ | $|E|$ | $\rho^*(G)$ | $\rho^*(G)/\tilde{\rho}(G)$ | | |
|---|---|---|---|---|---|---|
| | | | | $\epsilon = 0.001$ | $\epsilon = 0.1$ | $\epsilon = 1$ |
| as20000102 | 6,474 | 13,233 | 9.29 | 1.229 | 1.268 | 1.194 |
| ca-AstroPh | 18,772 | 396,160 | 32.12 | 1.147 | 1.156 | 1.273 |
| ca-CondMat | 23,133 | 186,936 | 13.47 | 1.072 | 1.072 | 1.429 |
| ca-GrQc | 5,242 | 28,980 | 22.39 | 1.000 | 1.000 | 1.395 |
| ca-HepPh | 12,008 | 237,010 | 119.00 | 1.000 | 1.017 | 1.151 |
| ca-HepTh | 9,877 | 51,971 | 15.50 | 1.000 | 1.000 | 1.356 |
| email-Enron | 36,692 | 367,662 | 37.34 | 1.058 | 1.072 | 1.063 |

Table 2: Empirical approximation bounds for various values of $\epsilon$.

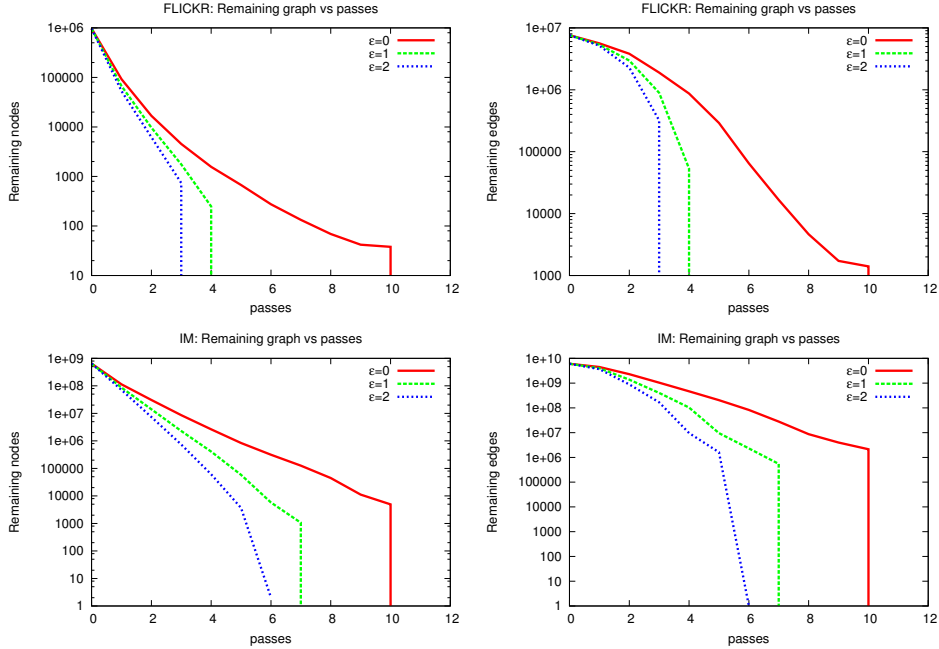

Figure 6.3: Number of nodes and edges in the graph after each step of the pass for FLICKR and IM.

1) and try $c$ at different powers of $\delta$ (One can prove that this worsens the approximation guarantee by at most a factor $\delta$ [10]). Clearly, the running time is given by $2 \log n / \log \delta$. First, we study the effect of the choice of $\delta$ compared to the choice of $\epsilon$. Table 3 shows the results. From the values, it is easy to see that as long as $\delta$ remains reasonable, the effect of $\epsilon$ is as in the undirected case. To make the rest of the study less cumbersome, we fix $\delta = 2$ for the remainder of this section. First, we present the results for LIVEJOURNAL.

| $\epsilon$ | $\delta$ | | |
|---|---|---|---|
| | 2 | 10 | 100 |
| 0 | 325.27 | 312.13 | 307.96 |
| 1 | 334.38 | 308.70 | 306.91 |
| 2 | 294.50 | 284.47 | 179.59 |

Table 3: LIVEJOURNAL: $\rho$ for different $\delta$ and $\epsilon$.

We study the performance of the algorithm for various choices of $c$, given $\delta = 2$. In particular, we measure the density and the number of passes. Figure 6.4 shows the values. The behavior of density is quite complex, and for LIVEJOURNAL, the optimum occurs when the relative sizes of $S$ and $T$ are not skewed.

Finally, Figure 6.5 shows the behavior of LIVEJOURNAL for the best setting of $c$ (which is 0.436) for $\delta = 2, \epsilon = 1$. It clearly shows the "alternate" nature of the simplified algorithm (Algorithm 3) that we developed in Section 4.3. As always, the number of nodes and edges fall dramatically as the passes progress.

For TWITTER, we used $\epsilon = 1, \delta = 2$ and studied the performance of the algorithm for various values of $c$. Figure 6.6 shows the density and the number of passes for various values of $c$. Unlike LIVEJOURNAL, the best value of $c$ is not concentrated around 1. This may be due to the highly skewed nature of the TWITTER graph: for example, there are about 600 popular users who are followed by more than 30 million other users. The results from LIVEJOURNAL and TWITTER suggest that, in practice, one can safely skip many values of $c$.

## 6.5 Performance of sketching

In this section we discuss the performance of the sketching heuristic presented in Section 5.1. We tested the algorithm on FLICKR, which has 976K nodes. Recall that the number of words in a Count-Sketch scheme using $b$ buckets and $t$ independent hash tables is $t \times b$. In Table 4, we show the ratio of the densest subgraph with and without sketching, for various values of $b$ and $\epsilon$. The bottom row shows the main memory used by the algorithm with sketching compared to the algorithm without sketching. Clearly, for small values of $\epsilon$, the performance difference is not very significant even for $b = 30000$, which means only $5 \times 30000/976K = 16\%$ of main memory is used. This suggests that, despite the space lower bounds (Lemma 7), in practice, a sketching scheme can obtain significant savings in main memory.

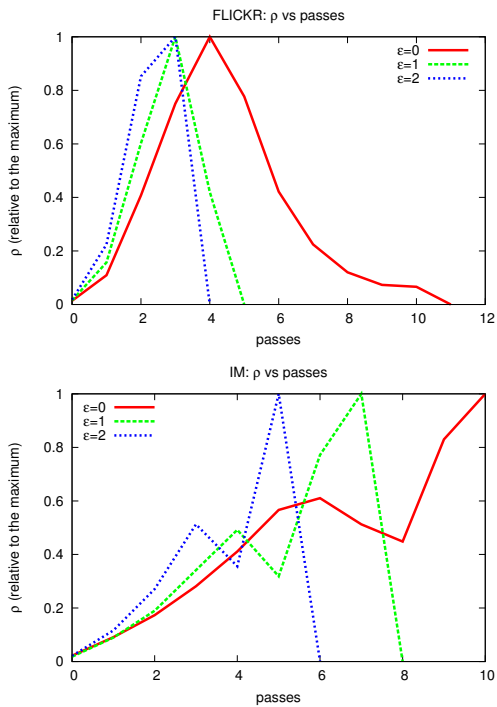

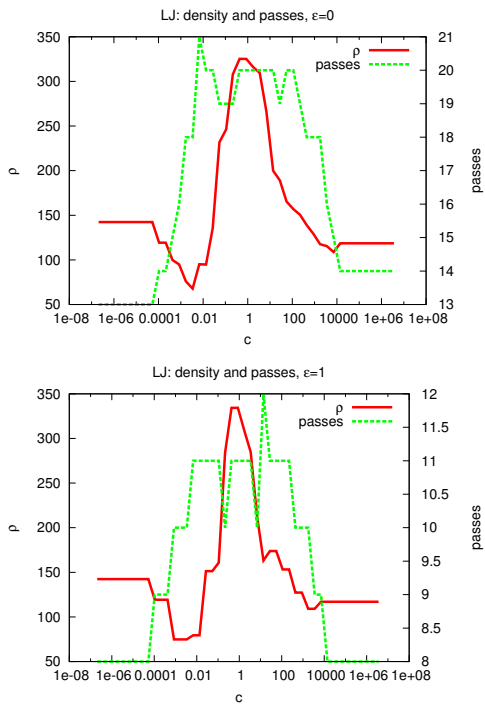

**Figure 6.2: Density as a function of the number of passes for various values of $\epsilon$, for FLICKR and IM.**

**Figure 6.4: Density and the number of passes at $\delta = 2$ for LIVEJOURNAL.**

| $\epsilon$ | $b = 30000$ | $b = 40000$ | $b = 50000$ |
|---|---|---|---|
| 0 | 1.047 | 1.027 | 1.014 |
| 0.5 | 0.960 | 0.896 | 0.921 |
| 1 | 0.958 | 0.936 | 0.918 |
| 1.5 | 0.890 | 0.911 | 0.929 |
| 2 | 0.760 | 0.845 | 0.869 |
| 2.5 | 0.787 | 0.708 | 0.740 |
| Memory | 0.16 | 0.20 | 0.25 |

**Table 4: Ratio of $\rho$ with and without sketching for FLICKR ($t = 5$).**

## 6.6 MapReduce implementation

In this section we study the performance of the MapReduce implementation of our algorithms for both directed and undirected graphs. For this purpose, we use the IM and TWITTER graphs since they are too big to be studied under the semi-streaming model. We implemented our algorithms in Hadoop (hadoop.apache.org) and ran it with 2000 mappers and 2000 reducers. Figure 6.7 shows the wall-clock running times for each pass for IM, which is an undirected graph. It only takes under 260 minutes for our algorithm to run on IM (a massive graph with more than half-billion nodes). For TWITTER, which is a directed graph, our algorithm takes around 35 minutes for a given value of $c$ and for each iteration; Figure 6.6 shows that the number of iterations is between four and seven, and the number of values of $c$ to be tried is very small. These clearly show the scalability of our algorithms.

## 7. CONCLUSIONS

In this paper we studied the problem of finding dense subgraphs, a fundamental primitive in several data management applications, in streaming and MapReduce, two computational models that are increasingly being adopted by large-scale data processing applications. We showed a simple algorithm that make a small number of passes over the graph and obtains a $(2+\epsilon)$-approximation to the densest subgraph. We then obtained several extensions of this algorithm: for the case when the the subgraph is prescribed to be more than a certain size and when the graph is directed. To the best of our knowledge, these are the first algorithms for the densest subgraph problem that truly scale yet offer provable guarantees. Our experiments showed that the algorithms are indeed scalable and achieve quality and performance that is often much better than the theoretical guarantees. Our algorithm's scalability is the main reason it was possible to run it on a graph with more than a half a billion nodes and six billion edges.

## Acknowledgments

We thank the anonymous reviewers for their many useful comments.

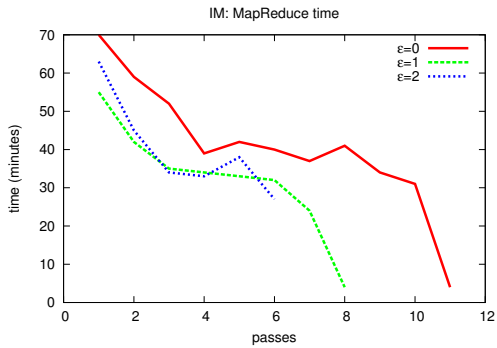

**Figure 6.7: Time taken on IM graph in MapReduce.**

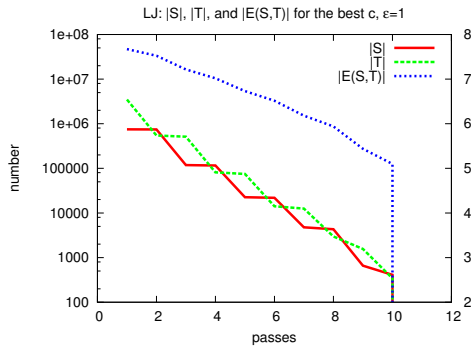

**Figure 6.5: Behavior of $|S|, |T|, |E(S,T)|$ for the best parameters of $c, \epsilon$ for LIVEJOURNAL.**

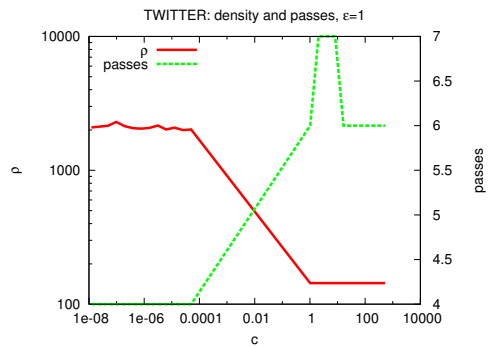

**Figure 6.6: Density and the number of passes at $\epsilon = 1, \delta = 2$ for TWITTER.**

## 8. REFERENCES

[1] C. C. Aggarwal and H. Wang. *Managing and Mining Graph Data*. Springer Publishing Company, Inc., 1st edition, 2010.
[2] R. Albert and A.-L. Barabási. Statistical mechanics of complex networks. *Rev. Mod. Phys.*, 74(1):47–97, 2002.
[3] R. Andersen and K. Chellapilla. Finding dense subgraphs with size bounds. In *WAW*, pages 25–37, 2009.
[4] Z. Bar-Yossef, T. S. Jayram, R. Kumar, and D. Sivakumar. An information statistics approach to data stream and communication complexity. *JCSS*, 68(4):702–732, 2004.
[5] Z. Bar-Yossef, R. Kumar, and D. Sivakumar. Reductions in streaming algorithms, with an application to counting triangles in graphs. In *SODA*, pages 623–632, 2002.
[6] L. Becchetti, P. Boldi, C. Castillo, and A. Gionis. Efficient semi-streaming algorithms for local triangle counting in massive graphs. In *KDD*, pages 16–24, 2008.
[7] B. Berger, J. Rompel, and P. W. Shor. Efficient NC algorithms for set cover with applications to learning and geometry. *JCSS*, 49(3):454–477, 1994.
[8] G. Buehrer and K. Chellapilla. A scalable pattern mining approach to web graph compression with communities. In *WSDM*, pages 95–106, 2008.
[9] A. Chakrabarti, S. Khot, and X. Sun. Near-optimal lower bounds on the multiparty communication complexity of set-disjointness. In *CCC*, pages 107–117, 2003.
[10] M. Charikar. Greedy approximation algorithms for finding dense components in a graph. In *APPROX*, pages 84–95, 2000.
[11] M. Charikar, K. Chen, and M. Farach-Colton. Finding frequent items in data streams. *TCS*, 312:3–15, 2004.
[12] J. Chen and Y. Saad. Dense subgraph extraction with application to community detection. *TKDE*, To appear.
[13] F. Chierichetti, R. Kumar, and A. Tomkins. Max-Cover in Map-Reduce. In *WWW*, pages 231–240, 2010.
[14] E. Cohen, E. Halperin, H. Kaplan, and U. Zwick. Reachability and distance queries via 2-hop labels. In *SODA*, pages 937–946, 2002.
[15] A. S. Das, M. Datar, A. Garg, and S. Rajaram. Google news personalization: scalable online collaborative filtering. In *WWW*, pages 271–280, 2007.
[16] J. Dean and S. Ghemawat. Mapreduce: simplified data processing on large clusters. In *OSDI*, pages 137–150, 2004.
[17] Y. Dourisboure, F. Geraci, and M. Pellegrini. Extraction and classification of dense communities in the web. In *WWW*, pages 461–470, 2007.
[18] J. Feigenbaum, S. Kannan, A. McGregor, S. Suri, and J. Zhang. On graph problems in a semi-streaming model. *TCS*, 348(2-3):207–216, 2005.
[19] J. Feldman, S. Muthukrishnan, A. Sidiropoulos, C. Stein, and Z. Svitkina. On distributing symmetric streaming computations. *TALG*, 6:1–19, 2010.
[20] A. Gajewar and A. D. Sarma. Multi-skill collaborative teams based on densest subgraphs. *CoRR*, abs/1102.3340, 2011.
[21] D. Gibson, R. Kumar, and A. Tomkins. Discovering large dense subgraphs in massive graphs. In *VLDB*, pages 721–732, 2005.
[22] A. V. Goldberg. Finding a maximum density subgraph. Technical Report UCB/CSD-84-171, EECS Department, University of California, Berkeley, 1984.
[23] R. Jin, Y. Xiang, N. Ruan, and D. Fuhry. 3-HOP: A high-compression indexing scheme for reachability query. In *SIGMOD*, pages 813–826, 2009.

[24] R. Kannan and V. Vinay. Analyzing the structure of large graphs, 1999. Manuscript.

[25] H. Karloff, S. Suri, and S. Vassilvitskii. A model of computation for mapreduce. In *SODA*, pages 938–948, 2010.

[26] S. Khuller and B. Saha. On finding dense subgraphs. In *ICALP*, pages 597–608, 2009.

[27] R. Kumar, J. Novak, and A. Tomkins. Structure and evolution of online social networks. In *KDD*, pages 611–617, 2006.

[28] S. Lattanzi, B. Moseley, S. Suri, and S. Vassilvitskii. Filtering: a method for solving graph problems in mapreduce. In *SPAA*, pages 85–94, 2011.

[29] E. Lawler. *Combinatorial Optimization: Networks and Matroids*. Holt, Rinehart, and Winston, 1976.

[30] J. Leskovec, K. J. Lang, A. Dasgupta, and M. W. Mahoney. Community structure in large networks: Natural cluster sizes and the absence of large well-defined clusters. *Internet Mathematics*, 6(1):29–123, 2009.

[31] A. McGregor. Finding graph matchings in data streams. In *APPROX*, pages 170–181, 2005.

[32] M. Newman. Modularity and community structure in networks. *PNAS*, 103(23):8577–8582, 2006.

[33] G. D. F. Morales, A. Gionis, and M. Sozio. Social content matching in mapreduce. *PVLDB*, pages 460–469, 2011.

[34] S. Muthukrishnan. Data streams: Algorithms and applications. *Foundations and Trends in Theoretical Computer Science*, 1(2), 2005.

[35] A. Nandi, C. Yu, P. Bohannon, and R. Ramakrishnan. Distributed cube materialization on holistic measures. In *ICDE*, pages 183–194, 2011.

[36] B. Saha, A. Hoch, S. Khuller, L. Raschid, and X.-N. Zhang. Dense subgraphs with restrictions and applications to gene annotation graphs. In *RECOMB*, pages 456–472, 2010.

[37] S. Suri and S. Vassilvitskii. Counting triangles and the curse of the last reducer. In *WWW*, pages 607–614, 2011.